# Sagging Ropes Demonstrate the Transversality Conditions of Variational Problems

## Am. J. Phys. 83 (12), 2015, pp. 998-1002.

Edward Bormashenko[a)], Gene Whyman, Yelena Bormashenko, Roman Grynyov.

Evgeny Shulzinger,

*Ariel University, Physics Department, Ariel, POB 3, 40700, Israel*

Alexander Kazachkov

*V. N. Karazin Kharkiv National University, School of Physics, Svobody Sq. 4, 61022,*

*Kharkiv, Ukraine.*

[a)]Corresponding author:

Edward Bormashenko

Ariel University, Physics Department,

P.O.B. 3, Ariel 40700, Israel

Phone: +972-3-906-6134

Fax: +972-3-906-6621

E-mail address: edward@ariel.ac.il

## Abstract

An account of the transversality conditions of variational problems gives rise to essential results in the analysis of different physical phenomena. This powerful and elegant approach has proven to be fruitful in a diversity of variational problems with free endpoints, when the endpoints are free to slip along preset curves.

We illustrate the transversality condition by the study of a heavy inextensible rope sagging both symmetrically and asymmetrically between two steering variously-shaped guide wires without friction. In this case, the transversality conditions lead to the orthogonality of the rope to the wires at endpoints of the rope, which is confirmed experimentally. Freeing the endpoints of the rope yields exact and simple analytical equations predicting the tension of the rope. Heavy ropes whose endpoints are free to slip between variously-shaped wires are discussed.

I. **INTRODUCTION**

Variational principles, formulated in the $17^{th}$-$18^{th}$ centuries by Fermat, Maupertuis and Euler, remain central in modern physics.[1-2] They serve as basic axiomatic foundations of mechanics (classical and quantum), field theory and even thermodynamics.[3-7] Variational principles present natural phenomena as problems of optimization under preset constraints. Typical variational problems involve the situation in which the physical quantity to be minimized (or maximized) appears as a stationary integral, i.e. a functional, because a function needs to be determined.[8] For example, to determine the shape of a chain deformed by gravity and suspended at both ends requires finding the coordinate-dependent function providing the minimum to the potential energy of a such chain. When the ends of a suspended inextensible chain are fixed, we have a typical variational problem, usually reduced to a solution of the Euler-Lagrange equation.[8]

In contrast, the variational problems with free endpoints, when the endpoints are free to slip along prescribed curves, are less known. These problems give rise to the so-called "transversality" conditions,[9] which turn out to be extremely instructive and fruitful in a broad class of physical problems, particularly cosmology, theories of wetting and elasticity.[10-12] For example, they enable prediction of contact angles of droplets placed on solid substrates (see Fig.1).[11] Moreover, they predict the contra-intuitive independence of contact angles on external fields, such as gravity or an electromagnetic field, which is validated in numerous experimental investigations.[13]

Our paper makes the transversality conditions understandable by illustrating them with a readily recognizable physical problem of a heavy rope sagging without friction between variously-shaped guide lines located in the same plane. Sagging ropes are ubiquitous in architecture, engineering, science, and arts. Examples of the use of the inverted catenary were found in Taq-i-Kisra in Ctesiphon (built approximately at 540 AD).[14] This makes the discussed problem attractive to the general reader.

## II. TRANSVERSALITY CONDITIONS OF VARIATIONAL PROBLEMS

Consider the functional

$$J(y) = \int_{x_o}^{x_1} F(x, y, y')dx \qquad (1)$$

defined on a smooth curve $y(x)$ whose ends, $x_0$ and $x_1$ are located on two given curves, $\varphi(x)$ and $\phi(x)$ (see Fig. 2). We seek to locate the extremum (which is the minimum in this case) of this functional by variation of the function $y(x)$. A typical problem of this kind is the calculation of the distance between two curves[9]. As it has been demonstrated (see e.g. Ref. 9), the function $y(x)$ supplying an extremum to the functional (1), whose limits are described by the curves $\varphi(x)$ and $\phi(x)$, has to satisfy the following boundary conditions

$$[F + (\phi' - y')F_{y'}]_{x=x_1} = 0, \qquad (2a)$$

$$[F + (\varphi' - y')F_{y'}]_{x=x_0} = 0, \qquad (2b)$$

where $F_{y'}$ denotes the $y'$ derivative of $F$. These boundary conditions are called the transversality conditions.[9] The function $y(x)$ satisfying Eqs. (2) is called "the transversal" to the functions $\varphi(x)$ and $\phi(x)$.[9]

In physics we often deal with functionals cast in the following form, and such is the functional describing sagging ropes or chains:

$$J(y) = \int_{x_o}^{x_1} f(x,y)\sqrt{1+y'^2}\,dx. \tag{3}$$

For these functionals, the transversality conditions appear in a very simple form (see Ref. 9):

$$y' = -\frac{1}{\varphi'}, \tag{4a}$$

$$y' = -\frac{1}{\phi'}. \tag{4b}$$

In other words, for the functionals given by Exp. (3), transversality means orthogonality.

### III. A CATENARY LINE WITH ENDPOINTS FREE TO SLIP ALONG STRAIGHT LINES.

The traditional problem of determining of the shape of a heavy, inextensible rope with the length $L$ and mass $M$, hanging symmetrically between fixed endpoints $x_1$ and $-x_1$ (see Fig. 3), is reduced to minimization of the functional

$$y_{cm} = \frac{2}{L}\int_0^{x_1} y\sqrt{1+y'^2}\,dx, \tag{5}$$

where $y_{cm}$ is the y-coordinate of the mass center of the rope. The rope will settle to the shape $y(x)$ for which the mass center will be in the lowest possible position, under the given constraints imposed by the constant length of the rope and its endpoints fixed. This shape is described by the *catenary* line

$$y(x) = a + \lambda\cosh\frac{x}{\lambda}, \tag{6}$$

where $\lambda = T_0/\mu g$; $T_0$ is the tension of the half-rope in its lowest point (see Fig. 3), and $\mu = M/L$ is the linear mass density of the rope (thus $\lambda$ is the tension $T_0$ normalized to the weight of the unit length). The constant $a$, appearing in Eq. (6), should be found from the condition: $y_1 = a + \lambda \cosh \frac{x_1}{\lambda}$, where $x_1, y_1$ are the fixed coordinates of the endpoints.

Now consider a heavy, inextensible rope sagging between ideally-smooth symmetrical guide pivots, as shown in Fig. 3. Its shape is described by the well-known equation of the catenary line, supplied by Eq. (6). This is due to the fact that the same functional (5) should be minimized; however, the endpoints of the rope are now *free to move* along the frictionless steering bars (pivots) defined by the equations: $\varphi(x) = -kx + b$, $\phi(x) = kx + b$; $k, b > 0$. The functional to be minimized belongs to the functionals described by Eq. (3) with $f(x, y) = y$. The transversality conditions (4a, 4b) express the orthogonality of the catenary line to the frictionless steering pivots at the endpoint $x_1$.

After the substitution of Exp. (6) into Condition (4a), we get

$$k \cdot \sinh \frac{x_1}{\lambda} = 1. \tag{7}$$

In turn, the constancy of the length $L$ of the rope gives rise to

$$L = 2\int_0^{x_1} \sqrt{1 + y'^2}\, dx = 2\int_0^{x_1} \sqrt{1 + \sinh^2 \frac{x}{\lambda}}\, dx = 2\int_0^{x_1} \cosh \frac{x}{\lambda}\, dx = 2\lambda \sinh \frac{x_1}{\lambda}. \tag{8}$$

Equations (7, 8) remarkably define not only the coordinates of the endpoints of the rope, but also the tension of the rope $T_0$. Indeed, dividing Eq. (8) by Eq. (7) leads to

$$L = \frac{2\lambda}{k} = \frac{2T_0}{\mu g k} \Rightarrow T_0 = \frac{1}{2} L\mu g k = \frac{1}{2} Mgk. \tag{9}$$

The *x*-coordinates of the endpoints of the rope should be calculated from

$$x_{1,2} = \pm \lambda \sinh^{-1}(1/k) = \pm \frac{T_0}{\mu g} \sinh^{-1}(1/k) = \pm \frac{1}{2} kL \sinh^{-1}(1/k). \tag{10}$$

This means that the endpoints of the rope will stop at the coordinates $x_{1,2}$ supplied by Eq. (10). A formal solution is given by Eq. (10) and Eq. (11):

$$\lambda = \frac{1}{2} kL. \tag{11}$$

Geometrically, the equilibrium shape of a rope with sliding ends may be characterized by its sagging defined as $h = y_1 - y_0$, where $y_0 = y(0)$, $y_1 = y(x_1)$ and by its half-width, $x_1$, appearing in Eq. (10) (see Fig. 4). For the present case of straight pivots, one gets from Eqs. (6), (10):

$$h = \frac{L}{2}\left(\sqrt{1+k^2} - k\right). \tag{12}$$

Dependence of the geometrical shape of the rope on the slope *k* of straight pivots is shown in Fig. 4. As is seen from Eqs. (10) and (12),

$$\lim_{k \to 0} x_1 = 0, \quad \lim_{k \to 0} h = \frac{L}{2}; \quad \lim_{k \to \infty} x_1 = \frac{L}{2}, \quad \lim_{k \to \infty} h = 0. \tag{13}$$

This means that under rotation of straight pivots, the rope folds vertically in two when they approach the horizontal position $(k \to 0)$; in the opposite limiting case, when the pivots approach the vertical position $(k \to \infty)$, the rope forms a horizontal segment.

Now consider in more detail the physics of the problem. When the endpoints of the rope are free to slide with a zero friction along guide bars, it is quite expectable that the rope at its endpoints will be normal to the pivots. Indeed, at rest and in the absence of friction, an elastic reaction of the pivots is normal to themselves, while a tension of the rope is tangential to the rope. Thus, the only scenario for the rope with free ends to stay at rest on the pivots is to be orthogonal to them in the contact (end) points (the orthogonality is obviously kept for a more general case of extensible ropes). The orthogonality is also expressed explicitly by the transversality conditions (4a, 4b) and (7). In order to check this prediction, we constructed an experimental unit, built of two straight strings (pivots) lubricated by silicone oil and supporting a heavy metallic chain shown in Fig. 5. We observed the orthogonality of the chain to the strings in all of our experiments, irrespective of the aperture angle (depicted in Fig. 5). The angle between the chain and pivots at the intersection points was established as $90 \pm 2.5^0$. It is noteworthy that the transversality conditions work also for asymmetrically sagging chains, as shown in Fig. 6.

For completeness, let us calculate the *y*-coordinate of the mass center of the sagging chain according to Eq. (5). Substitution of Eq. (6) into Eq. (5) yields:

$$y_{cm} = \frac{2a\lambda}{L}\sinh(x_1/\lambda) + \frac{\lambda^2}{L}\left(\frac{1}{2}\sinh(2x_1/\lambda) + \frac{x_1}{\lambda}\right) = \frac{\lambda}{L}((y_1 + a)\sinh(x_1/\lambda) + x_1). \quad (14)$$

Note that Eq. (14) is valid for a catenary line sliding along arbitrary symmetrical curves.

The position of the mass center of the rope on asymmetrical pivots $(x_1 \neq x_2, y_1 \neq y_2)$ is calculated as follows:

$$x_{cm} = \frac{\lambda}{L}(x_1\sinh(x_1/\lambda) - y_1 - x_2\sinh(x_2/\lambda) + y_2), \quad (15a)$$

$$y_{cm} = \frac{\lambda}{2L}((y_1 + a)\sinh(x_1/\lambda) + x_1 - (y_2 + a)\sinh(x_2/\lambda) - x_2). \quad (15b)$$

It is pertinent to note that the transversality conditions supplied by Eqs. 4a-b predict the orthogonality of the rope to pivots at its endpoints also for non-homogeneous ropes $\rho = \rho(x, y)$ sagging in a non-uniform field $g = g(x, y)$. In this case, the energy functional $E(y, y')$ to be minimized is given by

$$E(y, y') = \frac{1}{L} \int_{x_o}^{x_1} g(x, y) \rho(x, y) y \sqrt{1 + y'^2} \, dx. \tag{16}$$

It is readily seen that this expression is cast in form of Eq. (3); hence, the transversality conditions will be reduced to Eqs. (4a-b). Notice, however, that for the somewhat exotic field $g = g(x, y')$ this is not true.

Now consider contact angles of sessile water droplets, also resulting from imposing transversality conditions (2) on the variational problem of wetting (as mentioned in Introduction). The energy related to gravity depends on $y$ and $x$ but not on $y'$, and thus the more general transversality conditions (2) are not affected by gravity. As a result, contact angles of sessile droplets are independent of their masses.[11,13]

## IV. A CATENARY LINE WITH ENDPOINTS SLIDING ALONG VARIOUS CURVES

Consider the case of a heavy rope with endpoints slipping along two symmetric parabolas $\varphi(x) = \kappa(x-c)^2 + b$; $\phi(x) = \kappa(x+c)^2 + b$; $k, c, b > 0$. Applying Eqs. (4a), and (6) for the right endpoint, one gets

$$2\kappa(x_1 - c)\sinh\frac{x_1}{\lambda} = -1.$$

Then, after using Eq. (8),

$$x_1 = c - \frac{\lambda}{\kappa L} \tag{17}$$

and finally

$$2\frac{\lambda}{L}\sinh\left(\frac{c}{\lambda} - \frac{1}{\kappa L}\right) = 1. \tag{18}$$

Equation (18) determines the normalized tension $\lambda$ of the rope in its lowest point, and Eq. (17) gives the equilibrium coordinate $x_1$ of its sliding endpoint. The results of the numerical solution of Eq. (18) are presented in Figs. 7-8. As in the case of a linear pivot, the dimensionless tension in the apex increases with the increase of the pivot slope.

The expressions similar to (17), (18) describe the sagging of a heavy rope which is free to slide along guide lines having the shape of a catenary line: $y(x) = \alpha + \beta \cosh \frac{x \pm \gamma}{\beta}$ where α, β, γ are the parameters of the catenary guide lines. The transversality conditions (4a), (4b) yield

$$x_1 = \gamma - \beta \sinh^{-1}(2\lambda/L), \qquad (19)$$

$$\frac{2\lambda}{L} \sinh\left(\frac{\gamma - \beta \sinh^{-1}(2\lambda/L)}{\lambda}\right) = 1. \qquad (20)$$

The numerical solution of the system of equations (19), (20) may be accomplished analogously to that of Eq. (18), and it is exemplified in Fig. 9 by calculation of a shape of the rope. It is seen that the rope begins to fold when the catenary lines approach one another, and tends to a horizontal line as they separate from one another.

## V. DISCUSSION

Variational problems with free endpoints, when the endpoints are free to slip along prescribed curves or surfaces, are ubiquitous in physics. A typical problem of this kind is the variational problem of wetting, when a droplet is placed on a flat or curved surface[11]. In this case, the free energy of a droplet should be minimized in the situation where the contact line is free to slip along the solid surface.[11] Minimization of the functional representing the energy of such systems gives rise to the transversality conditions of the variational problem. These conditions comprise a rich physical content, exemplified in our paper by a simple mechanical system: a heavy chain which is free to slip in a frictionless way along straight or curved pivots. The transversality conditions predict orthogonality of the chain in its endpoints to the guide pivots. Moreover, they predict the chain tension (supplied by the explicit expression) depends on the slope of pivots. The orthogonality is preserved for non-homogenous ropes sagging in non-uniform fields, under broad assumptions about the nature of these fields. Experiments performed with chains sliding along straight and curved pivots validate the predictions supplied by the transversality conditions.

## APPENDIX: Constance of the horizontal component of rope tension

Consider the vertical component of the rope tension $T(x)$ at a point with the abscissa $x$, as depicted in Fig. 3. This component balances the weight of the half-rope with the mass $m(x)$, sagging between the origin of the coordinates and $x$. The equilibrium of the rope yields:

$$T(x)\sin\alpha = m(x)g. \qquad (21)$$

The mass $m(x)$ is supplied by

$$m(x) = \mu\int_0^x \sqrt{1+y'^2} = \lambda\mu\sinh(x/\lambda). \qquad (22)$$

The vertical component of the tension along the rope is thus given by

$$T(x)\sin\alpha(x) = \lambda\mu g\sinh(x/\lambda) = T_0\sinh(x/\lambda). \qquad (23)$$

The horizontal component of the rope tension $T(x)\cos\alpha$ is obtained immediately

$$T(x)\cos\alpha(x) = T_0\cot\alpha(x)\cdot\sinh(x/\lambda) \qquad (24)$$

Considering $\cot\alpha(x) = 1/y'(x) = 1/\sinh(x/\lambda)$ results in $T(x)\cos\alpha(x) = T_0$.

It also follows from Eq. (23) that

$$T(x) = \sqrt{T_0^2 + T^2(x)\sin^2\alpha(x)} = T_0\cosh(\mu g x/T_0). \qquad (25)$$

Thus, the tension is maximal at the maximal $x$ (at endpoints of the rope).

**References**

[1]C. Lanczos, *The Variational Principles of Mechanics*, (Dover Publications, NY, 1970), pp.5-6.

[2]J.-L Basdevant, *Variational Principles in Physics*, (Vuibert, Paris, France, 2007), pp.9-12.

[3]L.D. Landau, E.M. Lifshitz, *Mechanics. Vol. 1 (3rd ed.)*, (Butterworth-Heinemann, Oxford, 1976), pp. 2-4.

[4]L.D. Landau, E.M. Lifshitz, *The Classical Theory of Fields. Vol. 2 (4th ed.)*, (Butterworth-Heinemann, Oxford, 1975), pp. 24-25.

[5]R. P. Feynman, A. R Hibbs, *Quantum Mechanics and Path Integrals*, (McGraw-Hill, NY, 1965), pp. 26-31.

[6]L.M. Martyushev, V.D. Seleznev, "Maximum entropy production principle in physics, chemistry and biology", Physics Reports **426**, 1 – 45 (2006).

[7]I. Gyarmati, *Non-Equilibrium Thermodynamics: Field Theory and Variational Principle*, (Springer, NY, 1970), pp. 10-15.

[8]G. B. Arfken, H. J. Weber, *Mathematical Methods for Physicists (5$^{th}$. Ed)*, (Harcourt Academic Press, San Diego, 2001), pp. 1017-1052.

[9]I. M. Gelfand, S. V. Fomin, *Calculus of Variations*, (Dover Books on Mathematics, 2000), pp. 60-65.

[10]A. B. Whiting, "The Least-Action Principle: Theory of cosmological solutions and the radial velocity action", Astrophysical Journal, **533**, 50-61 (2000).

[11]E. Bormashenko, "Young, Boruvka–Neumann,Wenzel and Cassie–Baxter equations as the transversality conditions for the variational problem of wetting", Colloids & Surfaces A, **345**, 163–165 (2009).

[12]D. G. B. Edelen, "Aspects of variational arguments in the theory of elasticity: Fact and folklore", International Journal of Solids and Structures, **17**, 729–740 (1981).

[13]E. Bormashenko, *Wetting of Real Surfaces*, De Gruyter, Berlin, 2013.

[14]E. Conversano, M. Frangaviglia, Lorenzi M. G., L. Tedeschini-Lalli, "Persistence of form in art in architecture: catenaries, helicoids and sinusoids", J. Applied Mathematics, **4,** 101-112 (2011).

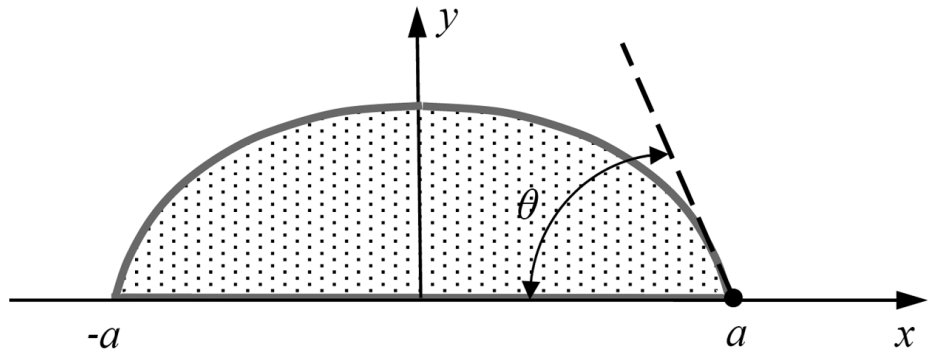

Fig. 1. The cross-section of an axisymmetric droplet placed on a solid surface. Endpoints $a$ and $-a$ are free to slip along a solid horizontal support. The angle $\theta$ formed between the tangent to the liquid surface and the solid support is called the "contact angle".

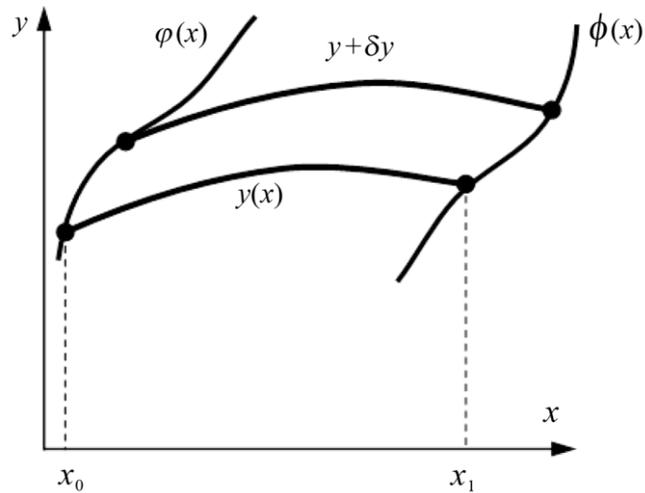

Fig. 2. Sketch illustrating the transversality conditions of the typical variational problem. Ends of the constant-length segment of the function $y(x)$ are free to slip along the curves $\varphi(x)$ and $\phi(x)$.

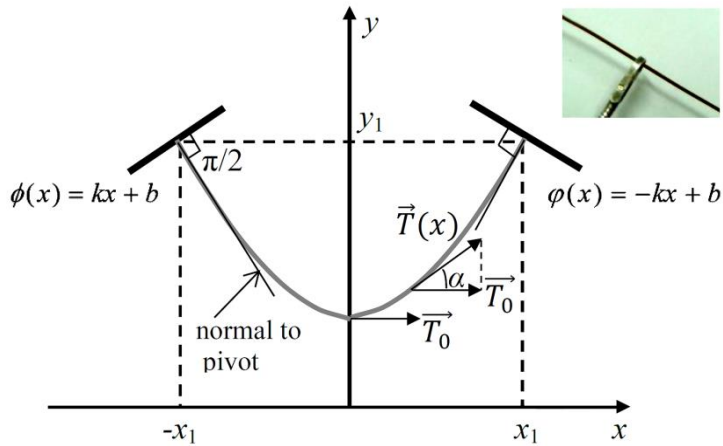

Fig. 3. Heavy rope (chain) sagging between straight frictionless pivots lying in the same vertical plane and described by: $\varphi(x) = -kx + b$; $\phi(x) = kx + b$; $k, b > 0$; $\vec{T}_0$ is the tension at the lowest point of the chain.

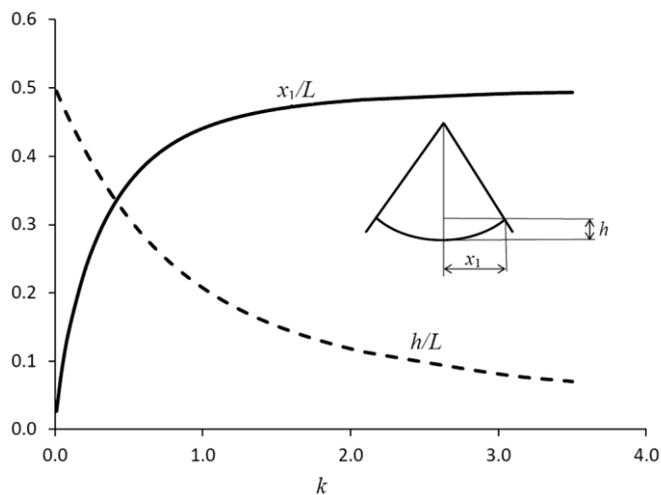

Fig. 4. The dimensionless sagging, $h/L$, and half-width, $x_1/L$, of the rope sliding along straight pivots as functions of the slope, $k$, of pivots.

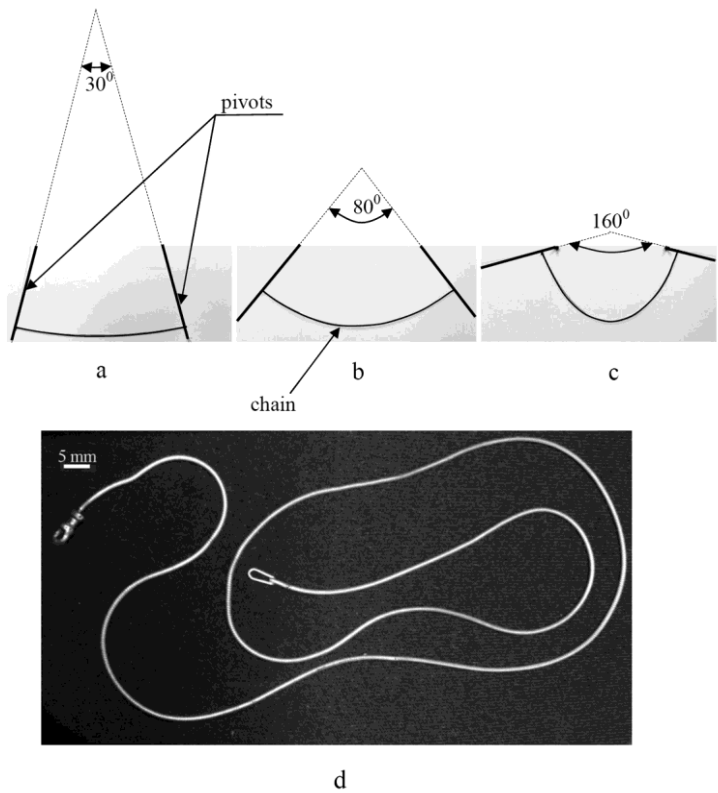

Fig. 5. a-c: Sequence of experimental images demonstrating a sagging of the chain suspended between lubricated pivots inclined symmetrically under various angles. d: The gold jewellery chain used in the investigation.

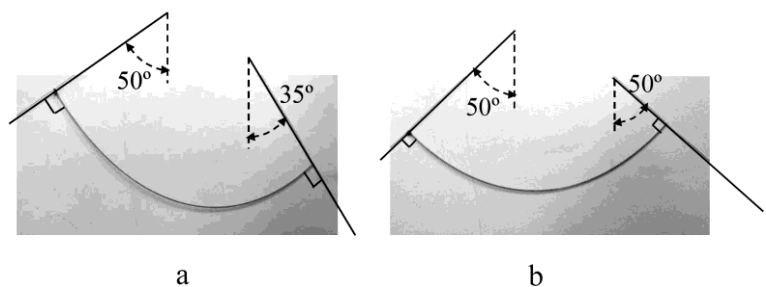

Fig. 6. Experimental. Asymmetrically sagging chain. a: Different pivot slopes and different heights of the suspension points of the pivots. b: Different heights of the suspension points of the pivots.

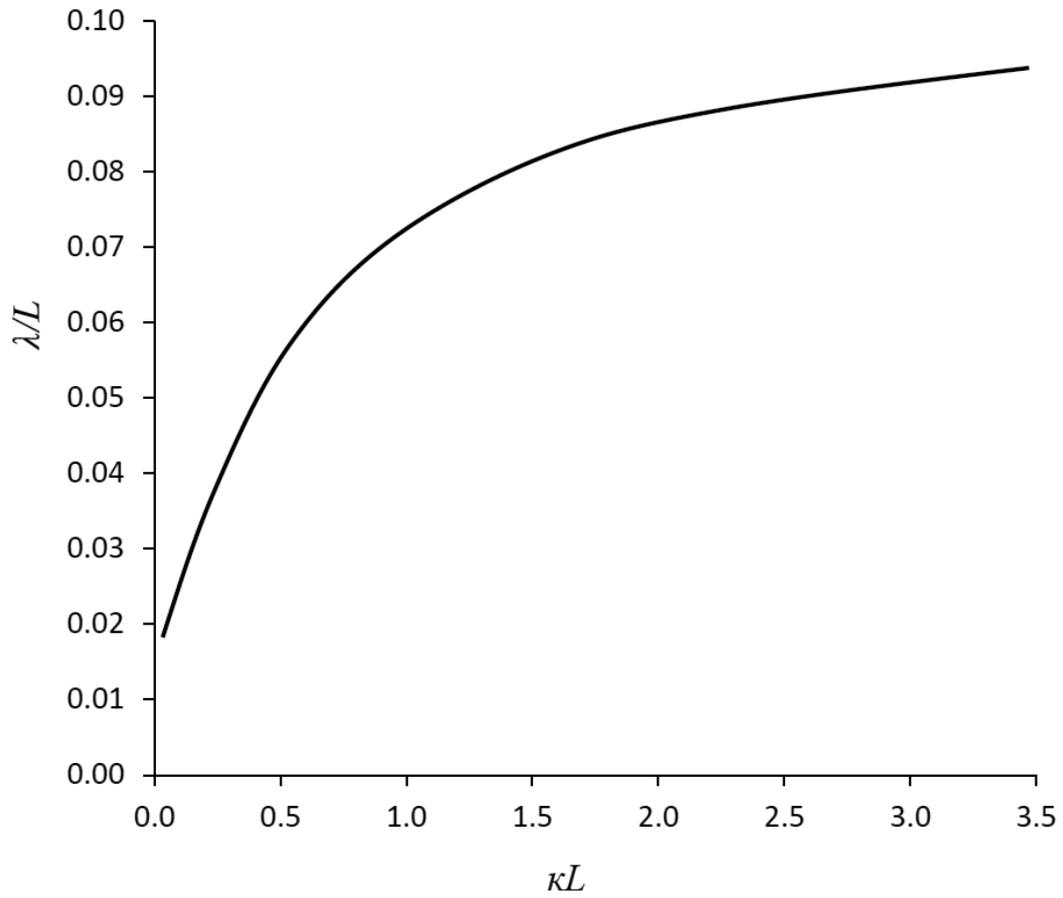

Fig. 7. The dimensionless tension, $\lambda/L$, at the rope apex vs. the dimensionless average slope, $\kappa L$, of the parabolic pivot described by $\varphi(x) = \kappa(x-c)^2 + b$; $c/L=1/4$.

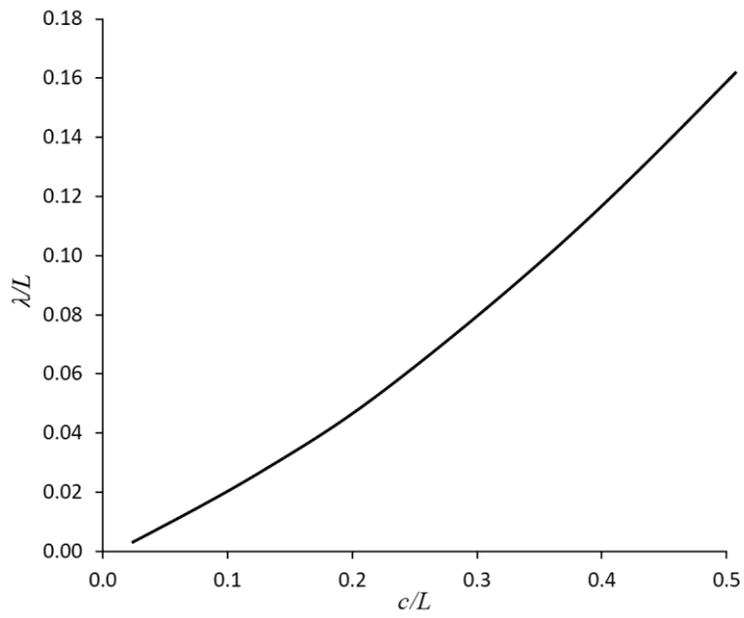

Fig. 8. The dimensionless tension, $\lambda/L$, at the rope apex vs. the dimensionless half-distance, $c/L$, between the apexes of the parabolic pivots; $\kappa L=5/6$.

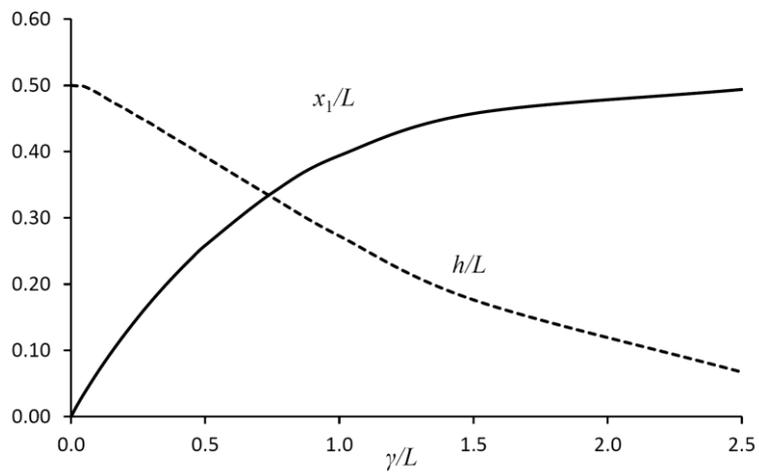

Fig. 9. Dependence of the sagging, $h$, and half-width, $x_1$, (see insert in Fig. 4) of the rope with endpoints sliding along two catenary lines on the half-distance, $\gamma$, between their vertices; $\beta/L = 1$.